\documentclass[fp,twocolumn]{jpsj3}

\title{Weyl Semimetallic State in the Rashba-Hubbard Model}

\author{Katsunori Kubo}
\inst{Advanced Science Research Center,
  Japan Atomic Energy Agency, Tokai, Ibaraki 319-1195, Japan}

\date{\today}

\abst{
We investigate the Hubbard model with the Rashba spin-orbit coupling
on a square lattice.
The Rashba spin-orbit coupling generates
two-dimensional Weyl points in the band dispersion.
In a system with edges along [11] direction, zero-energy edge states appear,
while no edge state exists for a system with edges along an axis direction.
The zero-energy edge states with a certain momentum along the edges
are predominantly in the up-spin state on the right edge,
while they are predominantly in the down-spin state on the left edge.
Thus, the zero-energy edge states are helical.
By using a variational Monte Carlo method for finite Coulomb interaction cases,
we find that the Weyl points can move toward the Fermi level
by the correlation effects.
We also investigate the magnetism of the model
by the Hartree-Fock approximation
and discuss weak magnetic order in the weak-coupling region.
}

\begin{document}
\maketitle

\section{Introduction}
In a two-dimensional system without inversion symmetry,
such as in an interface of a heterostructure,
a momentum-dependent spin-orbit coupling is allowed.
It is called the Rashba spin-orbit coupling~\cite{Bychkov1984}.
The Rashba spin-orbit coupling lifts the spin degeneracy
and affects the electronic state of materials.

Several interesting phenomena originating from the Rashba spin-orbit coupling
have been proposed and investigated.
%
By considering the spin precession by the Rashba spin-orbit coupling,
Datta and Das proposed the spin transistor~\cite{Datta1990},
in which electron transport between spin-polarized contacts
can be modulated by the gate voltage.
After this proposal,
the tunability of the Rashba spin-orbit coupling by the gate voltage
has been experimentally demonstrated~\cite{Schultz1996, Nitta1997, Engels1997}.
Such an effect may be used in a device in spintronics.
%
The possibility of the intrinsic spin Hall effect,
which is also important in the research field of spintronics,
by the Rashba spin-orbit coupling
has been discussed for a long time~\cite{Sinova2004, Inoue2004, Chalaev2005, Dimitrova2005, Sugimoto2006, Dugaev2010, Shitade2022}.
%
Another interesting phenomenon with the Rashba spin-orbit coupling
is superconductivity.
When the Rashba spin-orbit coupling is introduced in a superconducting system,
even- and odd-parity superconducting states are mixed
due to the breaking of the inversion symmetry~\cite{Gorkov2001, Yanase2008, Beyer2023}.
This mixing affects the magnetic properties of the superconducting state,
such as the Knight shift.

While the above studies have mainly focused on the one-electron states
in the presence of the Rashba spin-orbit coupling,
the effects of the Coulomb interaction between electrons
have also been investigated~\cite{Chen1999,Maryenko2021}.
The Hubbard model with the Rashba spin-orbit coupling on a square lattice
called the Rashba-Hubbard model
is one of the simplest models to investigate such effects.
In this study,
we investigate the ground state of this model at half-filling,
i.e., electron number per site $n=1$,
by the variational Monte Carlo method and the Hartree-Fock approximation.

In the strong coupling limit,
an effective localized model is derived
and the possibility of long-period magnetic order
is discussed~\cite{Cocks2012, Radic2012, Gong2015}.
The long-period magnetism is a consequence
of the Dzyaloshinskii-Moriya interaction
caused by the Rashba spin-orbit coupling.

Such magnetic order is also discussed by the Hartree-Fock approximation
for the Rashba-Hubbard model~\cite{Minar2013, Kennedy2022, Kawano2023}.
However, there is a contradiction among these studies
even within the Hartree-Fock approximation.
In the weak-coupling region with a finite Rashba spin-orbit coupling,
an antiferromagnetic order is obtained in Ref.~\citen{Minar2013},
but a paramagnetic phase is obtained in
Refs.~\citen{Kennedy2022} and \citen{Kawano2023}.
We will discuss this point in Sect.~\ref{sec: HF}.

The knowledge of the electron correlation beyond
the Hartree-Fock approximation is limited.
The electron correlation in the Rashba-Hubbard model
is studied by a dynamical mean-field theory
mainly focusing on magnetism~\cite{Zhang2015}
and by a cluster perturbation theory
investigating the Mott transition in the paramagnetic state~\cite{Brosco2020}.
We will study the electron correlation in the paramagnetic phase
by using the variational Monte Carlo method in Sect.~\ref{sec: VMC}.
The results concerning the Mott transition are consistent
with Ref.~\citen{Brosco2020}.
In addition, we find a transition to a Weyl semimetallic state
by the electron correlation.

Even without the Coulomb interaction,
the band structure of this model is intriguing.
When the Rashba spin-orbit coupling is finite,
the upper and lower bands touch each other at Weyl points.
In the large Rashba spin-orbit coupling limit,
all the Weyl points locate at the Fermi level for half-filling.
Topological aspects of the Weyl points
and corresponding edge states of this simple model
are discussed in Sect.~\ref{sec: topology}.

\section{Model}\label{model}
The model Hamiltonian is given by $H=H_{\text{kin}}+H_R+H_{\text{int}}$.
The kinetic energy term is given by
\begin{equation}
  H_{\text{kin}}
  =
  -t\sum_{(\mib{r},\mib{r}') \sigma}
  (c_{\mib{r} \sigma}^{\dagger}c_{\mib{r}' \sigma}
  +c_{\mib{r}' \sigma}^{\dagger}c_{\mib{r} \sigma})
  =\sum_{\mib{k} \sigma}
  \epsilon_{\mib{k}}
  c_{\mib{k} \sigma}^{\dagger}c_{\mib{k} \sigma},
\end{equation}
where $c_{\mib{r} \sigma}$ is the annihilation operator
of the electron at site $\mib{r}$ with spin $\sigma$
and $c_{\mib{k} \sigma}$ is the Fourier transform of it.
$(\mib{r},\mib{r}')$ denotes a pair of nearest-neighbor sites,
$t$ is the hopping integral,
and the kinetic energy is $\epsilon_{\mib{k}}=-2t (\cos k_x + \cos k_y)$,
where the lattice constant is set as unity.

The Rashba spin-orbit coupling term is given by~\cite{Mireles2001}
\begin{equation}
  \begin{split}
    H_R
    &=
    i\lambda_R
    \sum_{\mib{r} \sigma \sigma'  a=\pm 1}
    a
    \left(\sigma^x_{\sigma \sigma'}
    c_{\mib{r} \sigma}^{\dagger}c_{\mib{r}+a\hat{\mib{y}} \sigma'}
    -\sigma^y_{\sigma \sigma'}
    c_{\mib{r} \sigma}^{\dagger}c_{\mib{r}+a\hat{\mib{x}} \sigma'}
    \right)\\
    &=
    -2\lambda_R\sum_{\mib{k} \sigma \sigma'}
    \left(\sin k_y \sigma^x_{\sigma \sigma'}-\sin k_x \sigma^y_{\sigma \sigma'}
    \right)
    c_{\mib{k} \sigma}^{\dagger}c_{\mib{k} \sigma'}\\
    &=
    \sum_{\mib{k} \sigma \sigma'}
    \left[h_x(\mib{k}) \sigma^x_{\sigma \sigma'}+h_y(\mib{k}) \sigma^y_{\sigma \sigma'}
    \right]
    c_{\mib{k} \sigma}^{\dagger}c_{\mib{k} \sigma'}\\
    &=
    \sum_{\mib{k} \sigma \sigma'}
    H_{R \sigma \sigma'}(\mib{k})
    c_{\mib{k} \sigma}^{\dagger}c_{\mib{k} \sigma'},
  \end{split}
\end{equation}
where $\hat{\mib{x}}$ ($\hat{\mib{y}}$)
is the unit vector along the $x$ ($y$) direction,
$\mib{\sigma}$ are the Pauli matrices,
$\lambda_R$ is the coupling constant of the Rashba spin-orbit coupling,
$h_x(\mib{k})=-2\lambda_R \sin k_y$,
and
$h_y(\mib{k})= 2\lambda_R \sin k_x$.
We can assume $t \ge 0$ and $\lambda_R \ge 0$
without loss of generality.
We parametrize them as $t=\tilde{t}\cos \alpha$
and $\lambda_R=\sqrt{2} \, \tilde{t}\sin \alpha$.

\begin{figure}
  \includegraphics[width=0.99\linewidth]
    {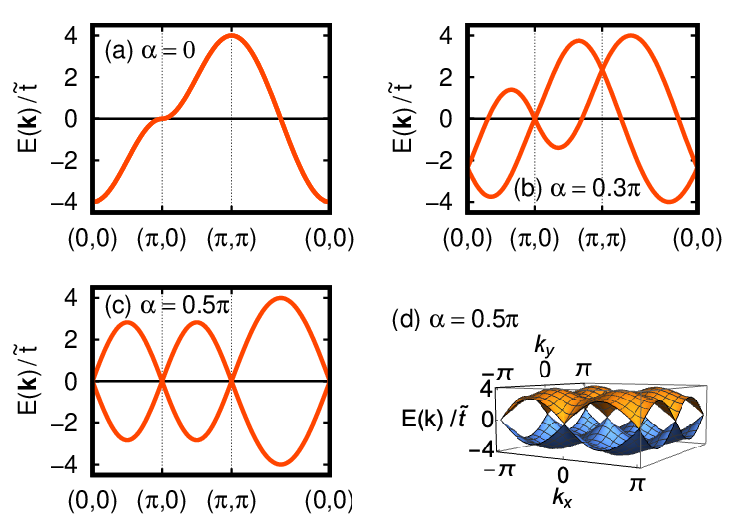}%
  \caption{(Color online)
    Energy dispersion along high symmetry directions
    for (a) $\alpha=0$, (b) $\alpha=0.3\pi$,
    and (c) $\alpha=0.5\pi$.
    (d) Energy dispersion in the entire Brillouin zone
    for $\alpha=0.5\pi$.
    \label{dispersion}}
\end{figure}

The band dispersion of $H_0=H_{\text{kin}}+H_R$ is
\begin{equation}
  E_{\pm}(\mib{k})
  =-2t(\cos k_x+\cos k_y) \pm |\mib{h}(\mib{k})|,
  \label{eq:dispersion}
\end{equation}
where $|\mib{h}(\mib{k})|=\sqrt{h_x^2(\mib{k})+h_y^2(\mib{k})}
=2\lambda_R\sqrt{\sin^2 k_x+\sin^2 k_y}$.
The bandwidth is $W=8\tilde{t}$.
Due to the electron-hole symmetry of the model,
the Fermi level is zero at half-filling.
For $\alpha=0$, that is, without the Rashba spin-orbit coupling,
the band is doubly degenerate [Fig.~\ref{dispersion}(a)].
For a finite $\lambda_R$, the spin degeneracy is lifted
except at the time-reversal invariant momenta
$\mib{X}^{(0)}=(0,0)$, $\mib{X}^{(1)}=(\pi,0)$, $\mib{X}^{(2)}=(0,\pi)$,
and $\mib{X}^{(3)}=(\pi,\pi)$ [Figs.~\ref{dispersion}(b) and \ref{dispersion}(c)].
These are two-dimensional Weyl points.
The energies at the Weyl points $\mib{X}^{(1)}$ and $\mib{X}^{(2)}$
are always zero.
By increasing $\alpha$ to $0.5\pi$ ($t=0$),
the energies at the other Weyl points
$\mib{X}^{(0)}$ and  $\mib{X}^{(3)}$ also move to zero.
In Fig.~\ref{dispersion}(d),
we show the energy dispersion in the entire Brillouin zone
for $\alpha=0.5\pi$.
We can see the linear dispersions around the Weyl points.

The Coulomb interaction term is given by
\begin{equation}
  H_{\text{int}}=U\sum_{\mib{r}}n_{\mib{r} \uparrow}n_{\mib{r} \downarrow},
\end{equation}
where $n_{\mib{r} \sigma}=c_{\mib{r} \sigma}^{\dagger}c_{\mib{r} \sigma}$
and $U$ is the coupling constant of the Coulomb interaction.

\section{Topology and edge states of the non-interacting Hamiltonian}\label{sec: topology}
The energy bands degenerate when $\mib{h}(\mib{k})=\mib{0}$,
i.e., at the Weyl points.
In the vicinity of these points, we set $\mib{k}=\mib{X}^{(l)}+\mib{p}$ and obtain
\begin{equation}
  \begin{split}
    H_R(\mib{k})
    &=
    \sum_j h_j(\mib{k})\sigma^j\\
    &\simeq
    \sum_{ij} \left.
    \frac{\partial h_j(\mib{k})}{\partial k_i} \right|_{\mib{k}=\mib{X}^{(l)}}
    p_i\sigma^j\\
    &=
    \sum_{ij}v^{(l)}_{ij}p_i\sigma^j.
  \end{split}
\end{equation}
The chirality of each Weyl point $\mib{X}^{(l)}$ is defined as
$\chi_l = \text{sgn} [ \det v^{(l)}]$~\cite{Hou2013}
and we obtain $\chi_0=\chi_3=1$ and $\chi_1=\chi_2=-1$.

\begin{figure}
  \includegraphics[width=0.99\linewidth]
    {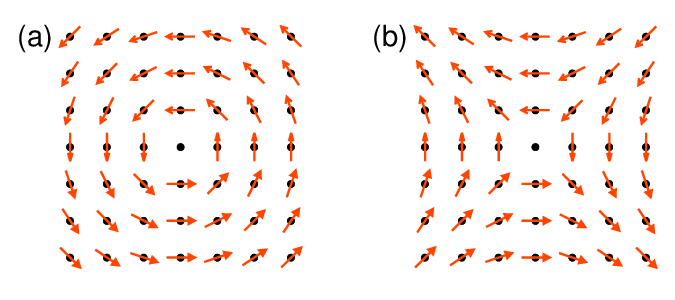}%
  \caption{(Color online)
    $\hat{\mib{h}}(\mib{k})$ (a) around $\mib{k}=\mib{X}^{(0)}=(0,0)$
    and (b) around $\mib{k}=\mib{X}^{(1)}=(\pi,0)$.
    \label{h_vector}}
\end{figure}

The winding number of a normalized two-component vector field
$\hat{\mib{h}}(\mib{k})=\mib{h}(\mib{k})/|\mib{h}(\mib{k})|$
is~\cite{Sun2012, Hou2013}
\begin{equation}
  w_l
  =
  \oint_{C_l} \frac{d\mib{k}}{2\pi}
  \cdot
  \left[
    \hat{h}_x(\mib{k})\mib{\nabla}\hat{h}_y(\mib{k})
    -\hat{h}_y(\mib{k})\mib{\nabla}\hat{h}_x(\mib{k})\right],
\end{equation}
where $C_l$ is a loop enclosing $\mib{X}_l$.
We obtain $w_l=\chi_l$.
Figure~\ref{h_vector} shows $\hat{\mib{h}}(\mib{k})$
around $\mib{k}=\mib{X}^{(0)}$ and $\mib{X}^{(1)}$ as examples.
We can recognize the winding numbers 1 and $-1$, respectively,
from this figure.

These topological numbers are related to the Berry phase~\cite{Berry1984}.
The eigenvector of $H_R(\mib{k})$ with eigenvalue $-|\mib{h}(\mib{k})|$ is
$|\mib{k}\rangle
=(1/\sqrt{2})(-1,\hat{h}_x(\mib{k})+i\hat{h}_y(\mib{k}))^{\text{T}}$.
The Berry connection is
\begin{equation}
  \begin{split}
    \mib{a}(\mib{k})
    &=
    i\langle \mib{k} | \mib{\nabla} |\mib{k}\rangle\\
    &=
    -\frac{1}{2}
    \left[
      \hat{h}_x(\mib{k})\mib{\nabla}\hat{h}_y(\mib{k})
      -\hat{h}_y(\mib{k})\mib{\nabla}\hat{h}_x(\mib{k})\right].
  \end{split}
\end{equation}
Then, the Berry phase is
\begin{equation}
  \gamma_l
  =
  \int_{C_l}d\mib{k} \cdot \mib{a}(\mib{k})
  =-w_l\pi.
\end{equation}

\begin{figure}
  \includegraphics[width=0.9\linewidth]
    {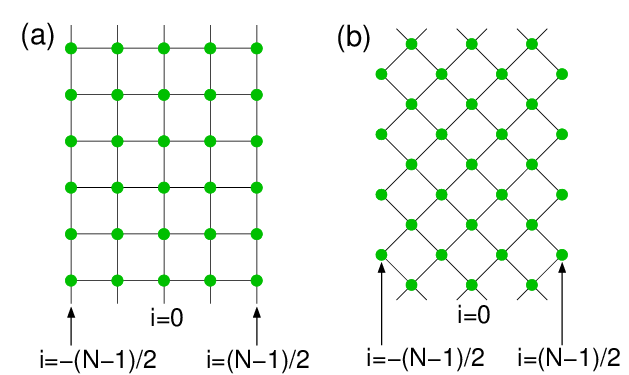}%
  \caption{(Color online)
    Lattices with edges.
    (a) Straight edges.
    (b) Zigzag edges.
    $i$ is the label denoting the position perpendicular to the edges,
    and $N$ is the number of these positions.
    \label{edge}}
\end{figure}

From the existence of such topological defects like the Weyl points,
we expect edge states as in graphene
with Dirac points~\cite{Fujita1996, Ryu2002, Hatsugai2009}.
We consider two types of edges:
the edges along an axis direction [straight edges, Fig.~\ref{edge}(a)]
and the edges along [11] direction [zigzag edges, Fig.~\ref{edge}(b)].
We denote the momentum along the edges as $k$
and the momentum perpendicular to the edges as $k_{\perp}$.
To discuss the existence of the edge states,
the chiral symmetry and
the winding number for a fixed $k$ are important~\cite{Ryu2002, Hatsugai2009}.
The Rashba term has a chiral symmetry:
$\{ H_R(\mib{k}), \sigma^z \} = H_R(\mib{k})\sigma^z+\sigma^zH_R(\mib{k})=0$
and $\sigma^z\sigma^{z \dagger}=I$ with $I$ being the unit matrix.
The winding number for a fixed $k$ is given by
\begin{equation}
  w(k)
  =
  \int_0^{2\pi} \frac{dk_{\perp}}{2\pi}
  \left[
  \hat{h}_x(\mib{k}) \frac{\partial}{\partial k_{\perp}} \hat{h}_y(\mib{k})
  -\hat{h}_y(\mib{k}) \frac{\partial}{\partial k_{\perp}} \hat{h}_x(\mib{k})
  \right].
\end{equation}
For the straight edges, we find $w(k)=0$
and we expect that the edge states are probably absent.
For the zigzag edges,
$h_x(\mib{k})=-2\lambda_R \sin(k-k_{\perp})$ and
$h_y(\mib{k})= 2\lambda_R \sin(k+k_{\perp})$,
where we have set $1/\sqrt{2}$ times the bond length as unity,
and we find $w(k)=-\text{sgn}[\sin(2k)]$
except for $k = 0$, $\pm\pi/2$, and $\pm\pi$ (projected Weyl points).
At the projected Weyl points, $w(k)=0$.
Thus, the edge states should exist except for the projected Weyl points
at least without $t$.

We note that the edge states can be understood
as those of a one-dimensional topological insulator.
The model only with the Rashba term with fixed $k$ is a one-dimensional model.
When this one-dimensional system has a gap
with a non-zero topological number,
the system can be regarded as a one-dimensional topological insulator
and has edge states.
This one-dimensional system is of symmetry class BDI
and can possess a topological number of $\mathbb{Z}$~\cite{Schnyder2008, Kitaev2009, Ryu2010}.

\begin{figure}
  \includegraphics[width=0.99\linewidth]
    {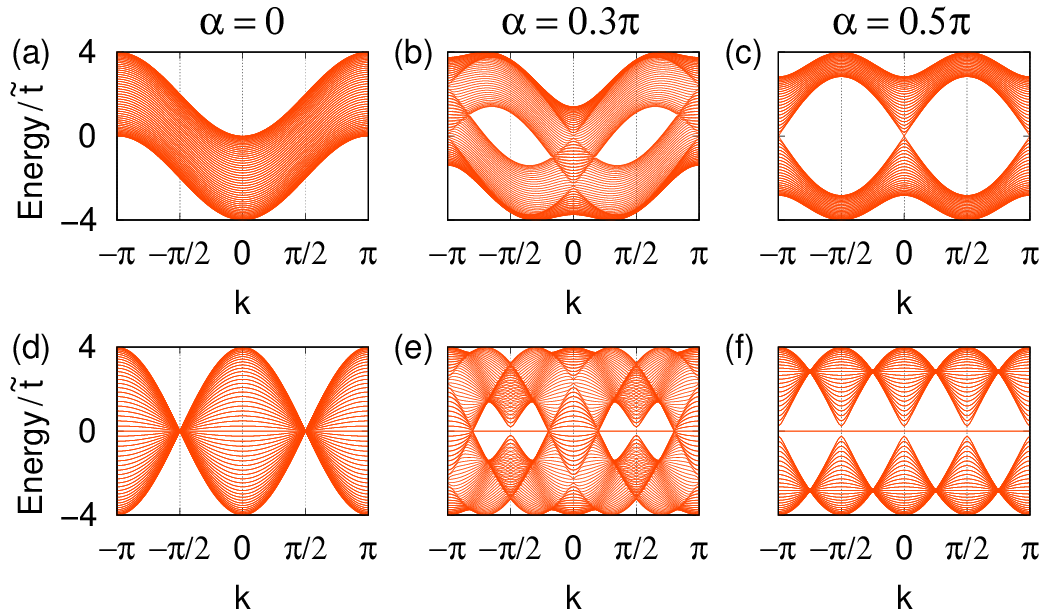}%
  \caption{(Color online)
    The upper panels show the band dispersion
    with straight edges:
    (a) for $\alpha=0$, (b) for $\alpha=0.3\pi$, and (c) for $\alpha=0.5\pi$.
    The lower panels show the band dispersion
    with zigzag edges:
    (d) for $\alpha=0$, (e) for $\alpha=0.3\pi$, and (f) for $\alpha=0.5\pi$.
    $k$ is the momentum parallel to the edges.
    The number of lattice sites perpendicular to the edges is $N=51$.
    \label{edge_band}}
\end{figure}

To explicitly demonstrate the existence of the edge states,
we numerically evaluate the band energy for lattices with finite widths.
We denote the number of lattice sites perpendicular to
the edges as $N$ (see Fig.~\ref{edge}) and obtain $2N$ bands.
The obtained energy bands are shown in Fig.~\ref{edge_band}.
For the straight edges [Figs.~\ref{edge_band}(a)--(c)],
we do not find the edge states.
It is consistent with $w(k)=0$.
For the zigzag edges [Figs.~\ref{edge_band}(d)--(f)],
we obtain isolated zero-energy states
except for $\lambda_R=0$ [Fig.~\ref{edge_band}(d)].
In particular, for $\alpha=0.5\pi$,
the zero-energy states appear at all the $k$ points
except for the projected Weyl points
as is expected from $w(k) \ne 0$.
We find that the zero-energy states remain even for finite $t$
as shown in Fig.~\ref{edge_band}(e).
For an even number of $N$,
the energy of the zero-energy states shifts from zero
around the projected Weyl points when $N$ is small.
For an odd number of $N$,
we obtain zero energy even for a small $N$.
Thus, we set $N=51$ in the calculations.

\begin{figure}
  \includegraphics[width=0.99\linewidth]
    {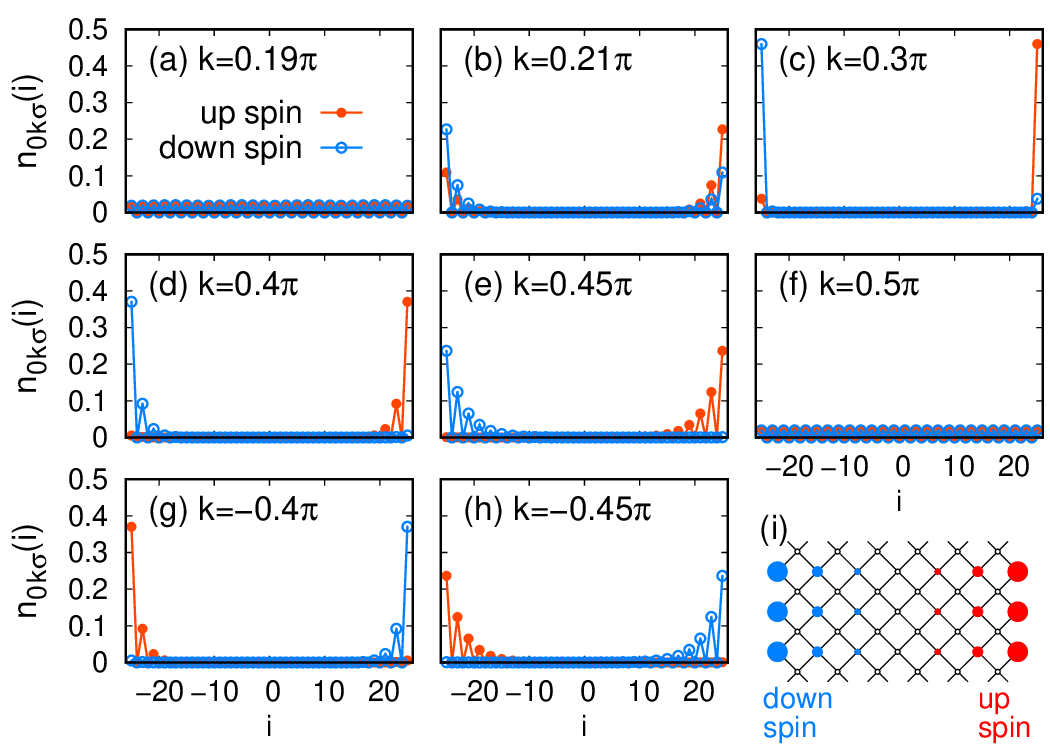}%
  \caption{(Color online)
    Spin density of the zero-energy edge states $n_{0 k \sigma}(i)$
    for $\alpha=0.3\pi$ on the lattice with zigzag edges with $N=51$:
    (a) for $k=0.19\pi$,
    (b) for $k=0.21\pi$,
    (c) for $k=0.3\pi$,
    (d) for $k=0.4\pi$,
    (e) for $k=0.45\pi$,
    (f) for $k=0.5\pi$,
    (g) for $k=-0.4\pi$,
    and (h) for $k=-0.45\pi$.
    (i) Schematic view of the spin density
    on the real-space lattice for $k \simeq 0.4\pi$.
    The sizes of the circles represent
    the magnitude of the spin density.
    \label{spin_density}}
\end{figure}

We discuss the characteristics of the zero-energy edge states.
We define $c_{i k\sigma}$ as the Fourier transform of $c_{\mib{r} \sigma}$
along the edges,
where $i$ labels the site perpendicular to the edges (see Fig.~\ref{edge}).
For the lattice with the zigzag edges,
we can show that
the states $c_{-(N-1)/2, \, \pi/4, \, \downarrow}^{\dagger}|0\rangle$
and
$c_{(N-1)/2, \, \pi/4, \, \uparrow}^{\dagger}|0\rangle$
do not have matrix elements of $H_R$,
where $|0\rangle$ is the vacuum state.
Thus, these states are the zero-energy states for $\alpha=0.5\pi$
completely localized on the left and right edges, respectively,
with opposite spins.
This helical character of the edge states is natural
since the system lacks inversion symmetry~\cite{Mella2022}
due to the Rashba spin-orbit coupling.
For other momenta and $\alpha$,
we calculate the spin density of the zero-energy edge states
$n_{0 k \sigma}(i)=\langle 0 k| c_{ik\sigma}^{\dagger} c_{ik\sigma}|0 k \rangle$,
where $|0 k \rangle$ denotes the zero-energy state at momentum $k$.
The zero-energy states are doubly degenerate,
and we take the average of the two states.
We show $n_{0 k \sigma}(i)$ for $\alpha=0.3\pi$, as an example,
in Fig.~\ref{spin_density}.
At $k$ where the bulk band gap is sufficiently large,
the zero-energy states are localized well on the edges
[Figs.~\ref{spin_density}(c) and \ref{spin_density}(d)].
As the bulk band gap becomes small,
the zero-energy states penetrate inner sites
[Figs.~\ref{spin_density}(b) and \ref{spin_density}(e)]
and the zero-energy states extend in the entire lattice when the gap closes
[Figs.~\ref{spin_density}(a) and \ref{spin_density}(f)].
The spin components are opposite between the edges.
For example, for $k=0.4\pi$ and $0.45\pi$,
the up-spin state dominates on the right edge
while the down-spin state dominates on the left edge.
Thus, the edge states are helical.
The spin components are exchanged between states at $k$ and $-k$
[compare Fig.~\ref{spin_density}(d) with Fig.~\ref{spin_density}(g)
  and Fig.~\ref{spin_density}(e) with Fig.~\ref{spin_density}(h)].
In Fig.~\ref{spin_density}(i),
we show a schematic view of the spin density corresponding to $k\simeq 0.4\pi$
on the real-space lattice.

\section{Weyl semimetallic state induced by the correlation effects}\label{sec: VMC}
In this section,
we investigate the effects of the Coulomb interaction $U$
at half-filling, i.e., the electron number per site $n=1$,
within the paramagnetic phase
by applying the variational Monte Carlo method~\cite{Yokoyama1987JPSJ56.1490}.
To achieve this objective,
it is necessary to select a wave function capable of describing
the Mott insulating state, as a Mott transition is anticipated,
at least in the ordinary Hubbard model without the Rashba spin-orbit coupling.
In this study,
we employ a wave function with doublon-holon binding factors
[doublon-holon binding wave function (DHWF)]~\cite{Kaplan1982, Yokoyama1990}.
A doublon means a doubly occupied site and a holon means an empty site.
Such intersite factors like doublon-holon binding factors are essential
to describe the Mott insulating state~\cite{Yokoyama2002, Capello2006}.
Indeed, the DHWF has succeeded in describing the Mott transition
for the single-orbital~\cite{Yokoyama2002, Watanabe2006, Yokoyama2006, Onari2007}
and two-orbital~\cite{Koga2006, Takenaka2012, Kubo2021, Kubo2022} Hubbard models.

The DHWF is given by
\begin{equation}
  |\Psi(\alpha_{\text{eff}})\rangle
  =
  P_d P_h P_G | \Phi(\alpha_{\text{eff}})\rangle.
\end{equation}
The Gutzwiller projection operator
\begin{equation}
  P_G=\prod_{\mib{r}}[1-(1-g)P_{d \, \mib{r}}],
\end{equation}
describes onsite correlations,
where
$P_{d \, \mib{r}} = n_{\mib{r} \uparrow}n_{\mib{r} \downarrow}$
is the projection operator onto the doublon state at $\mib{r}$
and $g$ is a variational parameter.
The parameter $g$ tunes the population of the doubly occupied sites.
When the onsite Coulomb interaction is strong and $n=1$,
most sites should be occupied by a single electron each.
In this situation,
if a doublon is created, a holon should be around it
to reduce the energy by using singly occupied virtual states.
$P_d$ and $P_h$ describe such doublon-holon binding effects.
$P_d$ is an operator to include intersite correlation effects
concerning the doublon states.
This is defined as follows~\cite{Kubo2021, Kubo2023, Kubo2022}:
\begin{equation}
  P_d=\prod_{\mib{r}}
  \left[1-(1-\zeta_d)
  P_{d \, \mib{r}}
  \prod_{\mib{a}}
  (1-P_{h \, \mib{r}+\mib{a}})
  \right],
\end{equation}
where
$P_{h \, \mib{r}} = (1-n_{\mib{r} \uparrow})(1-n_{\mib{r} \downarrow})$
is the projection operator onto the holon state at $\mib{r}$
and
$\mib{a}$ denotes the vectors connecting the nearest-neighbor sites.
$P_d$ gives factor $\zeta_{d}$
when site $\mib{r}$ is in the doublon state
and there is no holon at nearest-neighbor sites $\mib{r}+\mib{a}$.
Similarly,
$P_h$ describing the intersite correlation effects
on the holon state is defined as
\begin{equation}
  P_h=\prod_{\mib{r}}
  \left[1-(1-\zeta_h)
  P_{h \, \mib{r}}
  \prod_{\mib{a}}
  (1-P_{d \, \mib{r}+\mib{a}})
  \right].
\end{equation}
Factor $\zeta_h$ appears
when a holon exists without a nearest-neighboring doublon.
For the half-filled case,
we can use the relation $\zeta_d=\zeta_h$
due to the electron-hole symmetry of the model.

The one-electron part $|\Phi(\alpha_{\text{eff}}) \rangle$
of the wave function
is given by the ground state of the non-interacting Hamiltonian
$H_0(\alpha_{\text{eff}})$
in which $\alpha$ in $H_0$ is replaced by $\alpha_{\text{eff}}$.
We can choose $\alpha_{\text{eff}}$ different from
the original $\alpha$ in the model Hamiltonian.
Such a band renormalization effect of the one-electron part is discussed
for a Hubbard model with next-nearest-neighbor hopping~\cite{Sato2016}.
We define the normal state
as $|\Psi_{\text{N}}\rangle=|\Psi(\alpha_{\text{eff}}=\alpha)\rangle$,
i.e., $\alpha_{\text{eff}}$ remains the bare value.
We also define the Weyl semimetallic state
as $|\Psi_{\text{Weyl}}\rangle=|\Psi(\alpha_{\text{eff}}=0.5\pi)\rangle$,
i.e., all the Weyl points are at the Fermi level
and the Fermi surface disappears.
In addition, we can choose other values of $\alpha_{\text{eff}}$,
but in a finite-size lattice,
a slight change of $\alpha_{\text{eff}}$
does not change the set of the occupied wave numbers
and the wave function $|\Phi(\alpha_{\text{eff}}) \rangle$.
Thus, we have limited choices for $\alpha_{\text{eff}}$
as the band renormalization in the Hubbard model
with the next-nearest-neighbor hopping~\cite{Sato2016}.

We use the antiperiodic-periodic boundary conditions
since the closed shell condition is satisfied,
i.e., no $\mib{k}$ point is exactly on the Fermi surface
for a finite-size lattice
and there is no ambiguity to construct $|\Phi(\alpha_{\text{eff}})\rangle$.
The calculations are done for $L \times L$ lattices
with $L=12$, 14, and 16.

We evaluate the expectation value of energy by the Monte Carlo method.
We optimize the variational parameters $g$ and $\zeta_d=\zeta_h$
to minimize the energy.
We denote the optimized energy of $|\Psi(\alpha_{\text{eff}}) \rangle$
as $E(\alpha_{\text{eff}})$.
In particular, we denote $E_{\text{N}}=E(\alpha_{\text{eff}}=\alpha)$
and $E_{\text{Weyl}}=E(\alpha_{\text{eff}}=0.5\pi)$.
By using the Monte Carlo method,
we also evaluate the momentum distribution function
$n(\mib{k})=\sum_{\sigma} \langle c_{\mib{k} \sigma}^{\dagger}c_{\mib{k} \sigma} \rangle$,
where $\langle \cdots \rangle$ represents the expectation value
in the optimized wave function.

\begin{figure}
  \includegraphics[width=0.99\linewidth]
    {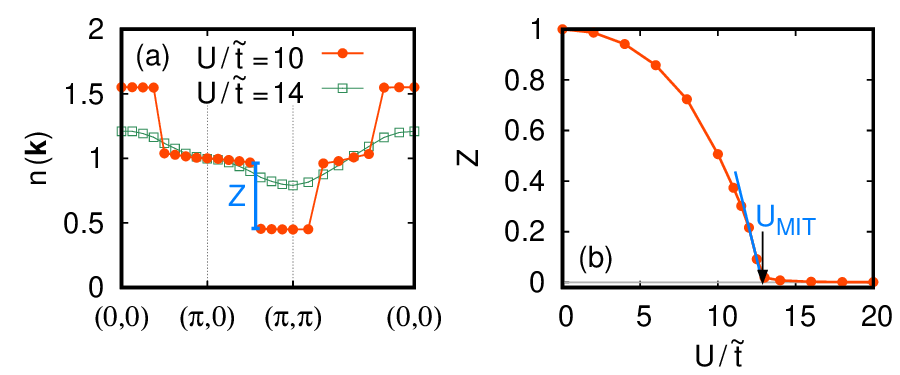}%
  \caption{(Color online)
    (a) Momentum distribution function $n(\mib{k})$
    for $\alpha=0.25\pi$ and $L=16$
    at $U=10\tilde{t}$ ($<U_{\text{MIT}}$)
    and $U=14\tilde{t}$ ($>U_{\text{MIT}}$).
    The quasiparticle renormalization factor $Z$
    is evaluated by the jump between $(\pi,0)$ and  $(\pi,\pi)$.
    Due to the antiperiodic boundary condition for the $x$ direction,
    we shift $k_x$ by $\pi/L$; for example,
    $(\pi,\pi)$ denoted in this figure actually means
    the point $(\pi-\pi/L,\pi)$.
    (b) Quasiparticle renormalization factor $Z$ as a function of $U$
    for $\alpha=0.25\pi$ and $L=16$.
    The Mott metal-insulator transition point $U_{\text{MIT}}$
    is determined by extrapolating the data to $Z = 0$.
    \label{nk_Z}}
\end{figure}

In Fig.~\ref{nk_Z}(a),
we show $n(\mib{k})$ in the normal state at $\alpha=0.25\pi$ for $L=16$.
For $U/\tilde{t}=10$,
$n(\mib{k})$ has clear discontinuities at the Fermi momenta.
On the other hand, for $U/\tilde{t}=14$,
$n(\mib{k})$ does not have such a discontinuity; that is,
the system is insulating
and a Mott metal-insulator transition takes place
between $U/\tilde{t}=10$ and  $U/\tilde{t}=14$.
To determine the Mott metal-insulator transition point $U_{\text{MIT}}$,
we evaluate the quasiparticle renormalization factor $Z$,
which is inversely proportional to the effective mass
and becomes zero in the Mott insulating state,
by the jump in $n(\mib{k})$.
Except for $\alpha=0$, we evaluate $Z$
by the jump between $(\pi,0)$ and $(\pi,\pi)$ as shown in Fig.~\ref{nk_Z}(a).
For $\alpha=0$, the above path does not intersect the Fermi surface
and we use the jump between $(\pi,\pi)$ and $(0,0)$ instead.
In Fig.~\ref{nk_Z}(b),
we show the $U$ dependence of $Z$ for $\alpha=0.25\pi$ and $L=16$.
By extrapolating $Z$ to zero,
we determine $U_{\text{MIT}}/\tilde{t} \simeq 12.9$.
We note that
for a small $\alpha$ with a large $L$,
the Mott transition becomes first-order
consistent with a previous study for $\alpha=0$~\cite{Yokoyama2006}.

\begin{figure}
  \includegraphics[width=0.99\linewidth]
    {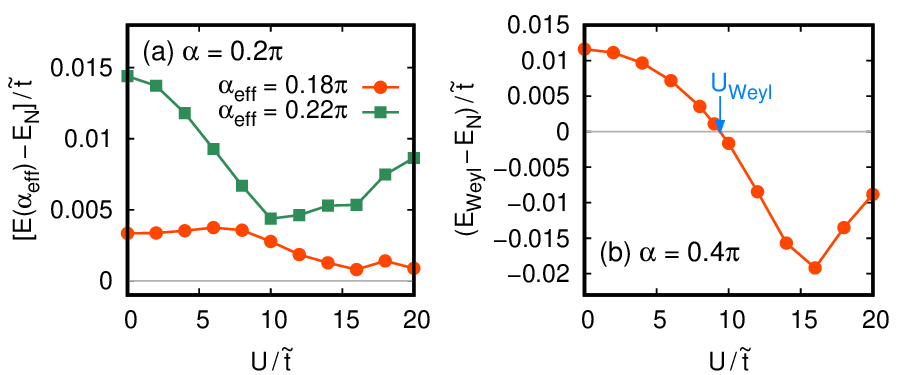}%
  \caption{(Color online)
    (a) Energy measured from that of the normal state
    [$E_{\text{N}}=E(\alpha_{\text{eff}}=\alpha)$]
    for $\alpha_{\text{eff}}=0.18\pi$ (circles)
    and $\alpha_{\text{eff}}=0.22\pi$ (squares)
    at $\alpha=0.2\pi$ for $L=16$.
    (b) Energy of the Weyl semimetallic state
    [$E_{\text{Weyl}}=E(\alpha_{\text{eff}}=0.5\pi)$]
    measured from that of the normal state
    at $\alpha=0.4\pi$ for $L=16$.
    The arrow indicates the Weyl transition point $U_{\text{Weyl}}$.
    \label{Energy}}
\end{figure}

We have also evaluated energies
for some values of $\alpha_{\text{eff}} \ne \alpha$.
Figure~\ref{Energy}(a) shows energies for $\alpha_{\text{eff}}=0.18\pi$
and $0.22\pi$ measured from the normal state energy at $\alpha=0.2\pi$
for $L=16$.
The normal state has the lowest energy, at least for $U/\tilde{t} \le 20$.
Thus, the renormalization of $\alpha$, even if it exists, is weak
for a system distant from the Weyl semimetallic state ($\alpha=0.5\pi$).
A similar conclusion is obtained
for a small intersite spin-orbit coupling case
of the Kane-Mele-Hubbard model~\cite{Richter2021}.
It is in contrast to the onsite spin-orbit coupling
case~\cite{Liu2023, Richter2021, Kubo2023, Kubo2022, Jiang2023},
where the effective spin-orbit coupling
is enhanced by the Coulomb interaction
even when the bare spin-orbit coupling is small.
On the other hand, the renormalization of $\alpha$ becomes strong
around $\alpha=0.5\pi$.
In Fig.~\ref{Energy}(b),
we show the energy $E_{\text{Weyl}}$ of the Weyl semimetallic state
measured from that of the normal state for $\alpha=0.4\pi$ for $L=16$.
$E_{\text{Weyl}}$ becomes lower than the normal state energy
at $U>U_{\text{Weyl}} \simeq 9.4\tilde{t}$.
There is a possibility that
the normal state changes to the Weyl semimetallic state gradually
by changing $\alpha_{\text{eff}}$ continuously.
However, for a finite lattice, the choices of $\alpha_{\text{eff}}$ are limited
between $\alpha_{\text{eff}}=\alpha$ and  $\alpha_{\text{eff}}=0.5\pi$.
For example, at $\alpha=0.4\pi$,
there is no choice for $L=12$ and $L=14$
and only one choice $0.4017<\alpha_{\text{eff}}/\pi<0.4559$ for $L=16$.
For this reason,
we evaluate $U_{\text{Weyl}}$ by comparing the energies
of the normal and the Weyl semimetallic states to show the tendency
toward the Weyl semimetallic state by the renormalization effect on $\alpha$.
Such a renormalization effect is also discussed
for a Dirac semimetal system~\cite{Fujioka2019}.

\begin{figure}
  \includegraphics[width=0.9\linewidth]
    {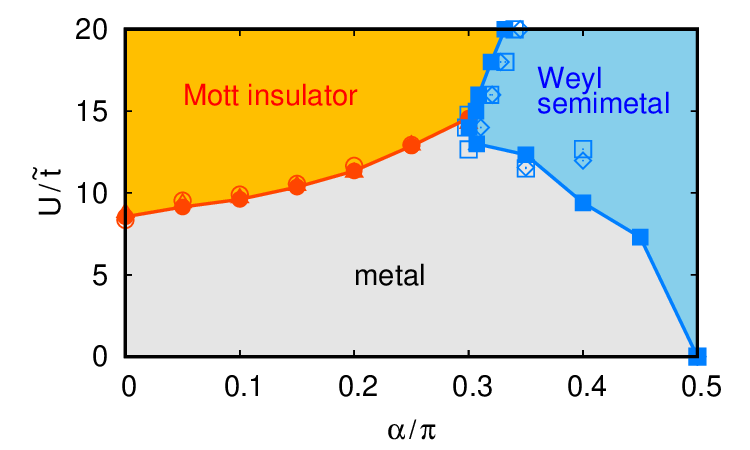}%
  \caption{(Color online)
    Phase diagram
    obtained with the variational Monte Carlo method
    without considering magnetic order.
    The Mott transition points for
    $L=12$ (open triangles),
    $L=14$ (open circles), and
    $L=16$ (solid circles)
    and
    the Weyl transition points for
    $L=12$ (open diamonds),
    $L=14$ (open squares), and
    $L=16$ (solid squares)
    are plotted.
    \label{VMC_phase_diagram}}
\end{figure}

Figure~\ref{VMC_phase_diagram} shows a phase diagram
without considering magnetic order.
The size dependence of the phase boundaries is weak.
For a weak Rashba spin-orbit coupling region,
i.e., for a small $\alpha$,
the Rashba spin-orbit coupling stabilizes the metallic phase.
It is consistent with a previous study
by a cluster perturbation theory~\cite{Brosco2020}.
Around $\alpha=0.5\pi$,
we obtain a wide region of the Weyl semimetallic phase.
Thus, we expect phenomena originating from the Weyl points
can be realized even away from $\alpha=0.5\pi$
with the aid of electron correlations.
In the Weyl semimetallic state,
the density of states at the Fermi level vanishes,
and thus, energy gain is expected
similar to the energy gain by a gap opening
in an antiferromagnetic transition.
We note that such a renormalization effect on $\alpha$
cannot be expected within the Hartree-Fock approximation
and is a result of the electron correlations
beyond the Hartree-Fock approximation.

\section{Hartree-Fock approximation for magnetism}\label{sec: HF}
\begin{figure}
  \includegraphics[width=0.99\linewidth]
    {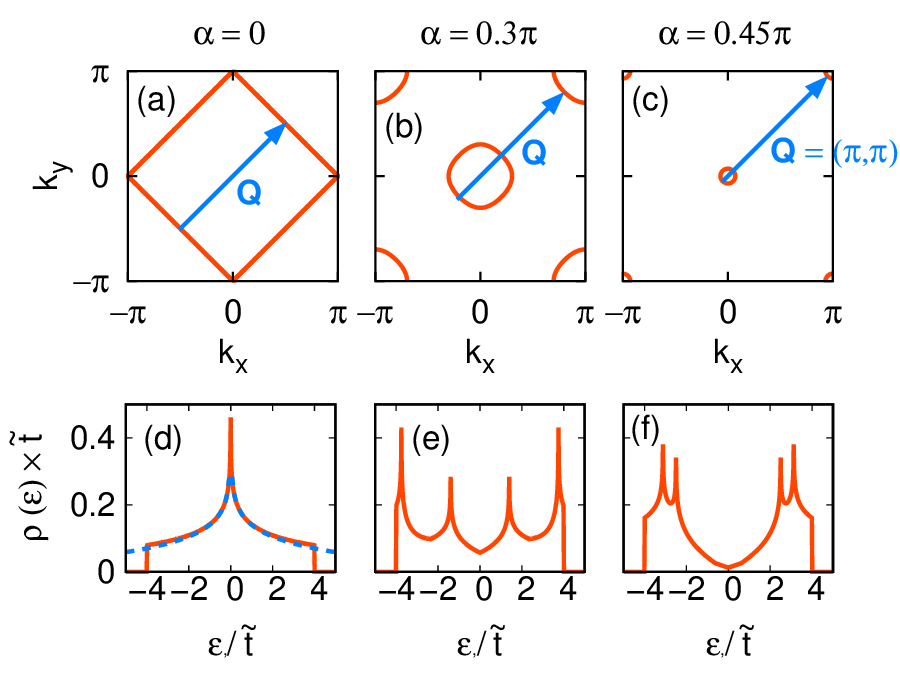}%
  \caption{(Color online)
    Fermi surface
    (a) for $\alpha=0$,
    (b) for $\alpha=0.3\pi$, and
    (c) for $\alpha=0.45\pi$.
    The arrows represent the nesting vector $\mib{Q}=(\pi,\pi)$.
    Density of states
    (d) for $\alpha=0$,
    (e) for $\alpha=0.3\pi$, and
    (f) for $\alpha=0.45\pi$.
    The dashed line in (d) is the asymptotic form
    $-[1/(2\pi^2\tilde{t})]\ln [|\epsilon|/(16\tilde{t})]$.
    \label{FS_DOS}}
\end{figure}

In this section, we discuss the magnetism of the model
by the Hartree-Fock approximation.
The energy dispersion given in Eq.~\eqref{eq:dispersion}
has the following property:
$E_{\pm}(\mib{k}+\mib{Q})=-E_{\mp}(\mib{k})$ for $\mib{Q}=(\pi,\pi)$.
When $E_a(\mib{k})=0$, in particular,
$E_{-a}(\mib{k}+\mib{Q})=E_a(\mib{k})=0$.
Thus, the Fermi surface is perfectly nested
for half-filling (the Fermi energy is zero)
with the nesting vector $\mib{Q}=(\pi,\pi)$ [see Figs.~\ref{FS_DOS}(a)--(c)].
Due to this nesting,
the magnetic susceptibility at $\mib{Q}=(\pi,\pi)$
of the non-interacting system
diverges at zero temperature~\cite{Minar2013}.
It indicates that the magnetic order occurs
with an infinitesimally small value of the Coulomb interaction $U$
at zero temperature.
However, some recent Hartree-Fock studies argue
the existence of the paramagnetic phase
with finite $U$~\cite{Kennedy2022, Kawano2023}.
To resolve this contradiction
and gain insights into magnetism,
we apply the Hartree-Fock approximation to the model
within two-sublattice magnetic order,
i.e., with ordering vector of $\mib{Q}=(\pi,\pi)$ or $\mib{Q}=(\pi,0)$.

The Hartree-Fock Hamiltonian is given by
\begin{equation}
  H_{\text{HF}}
  =
  \sum_{\mib{k}}
  \begin{pmatrix}
    c_{\mib{k}}^{\dagger}
    c_{\mib{k}+\mib{Q}}^{\dagger}
  \end{pmatrix}
  \begin{pmatrix}
    \hat{\epsilon}(\mib{k}) & -\mib{\Delta}\cdot\mib{\sigma} \\
    -\mib{\Delta}\cdot\mib{\sigma} & \hat{\epsilon}(\mib{k}+\mib{Q})
  \end{pmatrix}
  \begin{pmatrix}
    c_{\mib{k}} \\
    c_{\mib{k}+\mib{Q}}
  \end{pmatrix},
\end{equation}
where
$\mib{k}$-summation runs over the folded Brillouin zone
of the antiferromagnetic state,
$c_{\mib{k}}=(c_{\mib{k} \uparrow},c_{\mib{k} \downarrow})^{\text{T}}$,
$\hat{\epsilon}(\mib{k})=\epsilon_{\mib{k}}I+H_R(\mib{k})$,
and $\mib{\Delta}=U\mib{m}_{\text{AF}}$.
Here, $\mib{m}_{\text{AF}}=[1/(2L^2)]\sum_{\mib{r} \sigma \sigma'}e^{-i\mib{Q}\cdot\mib{r}}
\langle
c_{\mib{r} \sigma}^{\dagger} \mib{\sigma}_{\sigma \sigma'} c_{\mib{r} \sigma'}
\rangle_{\text{HF}}$,
where $\langle \cdots \rangle_{\text{HF}}$ represents the expectation value
in the ground state of $H_{\text{HF}}$.
We solve the gap equation $\mib{\Delta}=U\mib{m}_{\text{AF}}$ self-consistently.

\begin{figure}
  \includegraphics[width=0.99\linewidth]
    {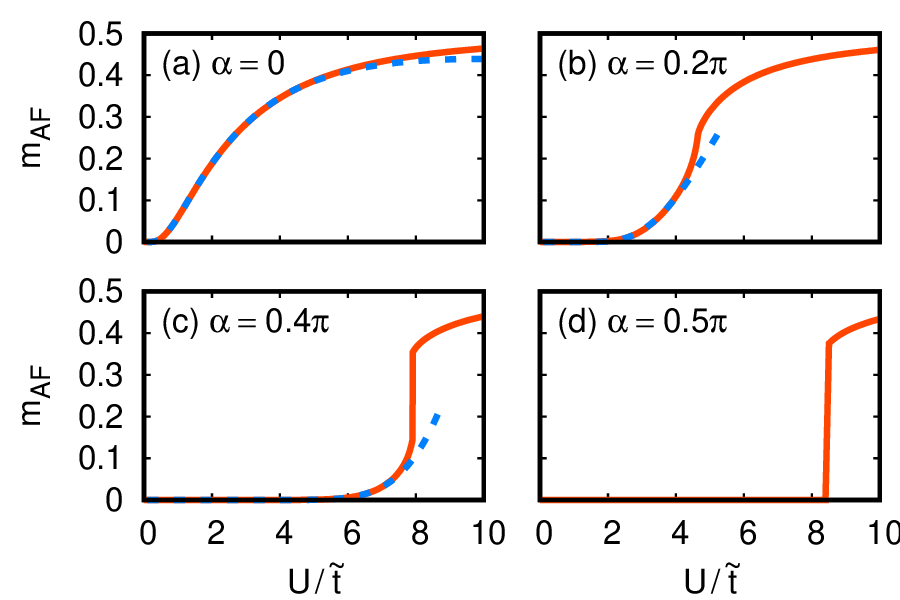}%
  \caption{(Color online)
    Antiferromagnetic moment $m_{\text{AF}}$ with $\mib{Q}=(\pi,\pi)$
    for (a) $\alpha=0$, (b) $\alpha=0.2\pi$, (c) $\alpha=0.4\pi$,
    and (d) $\alpha=0.5\pi$.
    The dashed line in (a) is $(32\tilde{t}/U)e^{-2\pi\sqrt{\tilde{t}/U}}$.
    The dashed lines in (b) and (c)
    are fitted curves with $(a\tilde{t}/U)e^{-2/[U\rho(0)]}$.
    We used an $L \times L$ lattice with $L=200$
    except for $\alpha=0$ with $U<\tilde{t}$.
    For $\alpha=0$ with $U<\tilde{t}$,
    the finite size effect is severe
    and we extrapolated the magnetization for $L \rightarrow \infty$
    by using data of $L=400$, 1000, and 2000.
    \label{m_AF}}
\end{figure}

First, we consider the magnetic order for $\mib{Q}=(\pi,\pi)$.
Without the Rashba spin-orbit coupling,
the asymptotic form
$m_{\text{AF}}=|\mib{m}_{\text{AF}}|\sim (\tilde{t}/U)e^{-2\pi\sqrt{\tilde{t}/U}}$
for the weak-coupling region $\Delta=|\mib{\Delta}| \ll W$
was obtained by Hirsch analyzing the gap equation~\cite{Hirsch1985}.
If we take into consideration the fact
that the asymptotic form of the density of states
$\rho(\epsilon) \simeq -[1/(2\pi^2 \tilde{t})] \ln [|\epsilon|/(16\tilde{t})]$
for $\epsilon \simeq 0$~\cite{Fazekas1999}
is a good approximation even up to the band edge [see Fig.~\ref{FS_DOS}(d)],
we obtain $m \simeq (32\tilde{t}/U)e^{-2\pi\sqrt{\tilde{t}/U}}$.
Indeed, this approximate form reproduces the numerical data well
in the weak-coupling region as shown in Fig.~\ref{m_AF}(a).

For a finite $\lambda_R$,
we find numerically that
$\mib{m}_{\text{AF}}$ is parallel to the $x$ or $y$ direction.
It is expected from the effective Hamiltonian
in the strong coupling limit we will discuss later.
By assuming $\Delta \ll \lambda_R$ and $\Delta \ll W$,
we obtain $m_{\text{AF}} \sim (\tilde{t}/U)e^{-2/[U\rho(0)]}$ for a finite $\rho(0)$,
where $\rho(0)$ is the density of states at the Fermi level.
The coefficient to $m_{\text{AF}}$ is determined by the entire behavior
of the density of states up to the band edge
[see Figs.~\ref{FS_DOS}(e) and \ref{FS_DOS}(f)]
and we cannot obtain it analytically in general.
Figures~\ref{m_AF}(b) and \ref{m_AF}(c) show the numerically obtained
$m_{\text{AF}}$ for $\alpha=0.2\pi$ and $0.4\pi$, respectively,
along with the fitted curves of $(a\tilde{t}/U)e^{-2/[U\rho(0)]}$,
where $a$ is the fitting parameter.
The fitted curves reproduce well the numerical data
in the weak-coupling region.

From the obtained asymptotic form and the numerical data supporting it,
we conclude that the magnetic order occurs
by an infinitesimally small $U$ for $0 \le \alpha < 0.5\pi$
consistent with the divergence of the magnetic susceptibility~\cite{Minar2013}.
We cannot apply this asymptotic form for $\alpha=0.5\pi$
since $\rho(0)=0$ there.
The numerical result shown in Fig.~\ref{m_AF}(d) indicates a first-order transition
for $\alpha=0.5\pi$.

Note that it is possible to include magnetic order
in the variational Monte Carlo calculations~\cite{Yokoyama1987JPSJ56.3582, Kubo2009JPCS, Kubo2009PRB, Kubo2011JPSJSA, Kubo2015JPCS, Kubo2015JPSJ, Kubo2015PP, Kubo2017JPSJ, Kubo2021, Kubo2022}.
We have also performed such variational Monte Carlo calculations
incorporating magnetic order for the present model
and observed the emergence of weak magnetic order for small values of $U$,
consistent with the results obtained via the Hartree-Fock approximation.

Here, we discuss previous papers indicating
the existence of the paramagnetic phase with finite $U$.
In Ref.~\citen{Kennedy2022},
the authors introduced a threshold $\varepsilon$
for the magnetization $m_{\text{AF}}$.
Then, the authors determined the magnetic transition point
when $m_{\text{AF}}$ becomes smaller than $\varepsilon$.
However, $m_{\text{AF}}$ becomes exponentially small in the weak-coupling region,
as understood from the above analysis.
In Ref.~\citen{Kennedy2022},
$\varepsilon$ is not sufficiently small
to discuss the exponentially small value of $m_{\text{AF}}$
and a finite region of the paramagnetic phase was obtained.
In Ref.~\citen{Kawano2023},
the authors calculated the energy difference $\Delta E$
between the paramagnetic state and the antiferromagnetic state.
Then, the authors introduced a scaling
between $\Delta E$ and $U-U_{\text{AF}}$,
where $U_{\text{AF}}$ is the antiferromagnetic transition point.
They tuned $U_{\text{AF}}$ to collapse the data with different $\alpha$
onto a single curve in a large-$U$ region.
Then, they obtained finite $U_{\text{AF}}$ for $\alpha \ne 0$.
However, this scaling analysis does not have a basis.
In particular, if such a scaling holds for critical behavior,
the data collapse should occur for $U \simeq U_{\text{AF}}$,
not for a large-$U$ region.

\begin{figure}
  \includegraphics[width=0.9\linewidth]
    {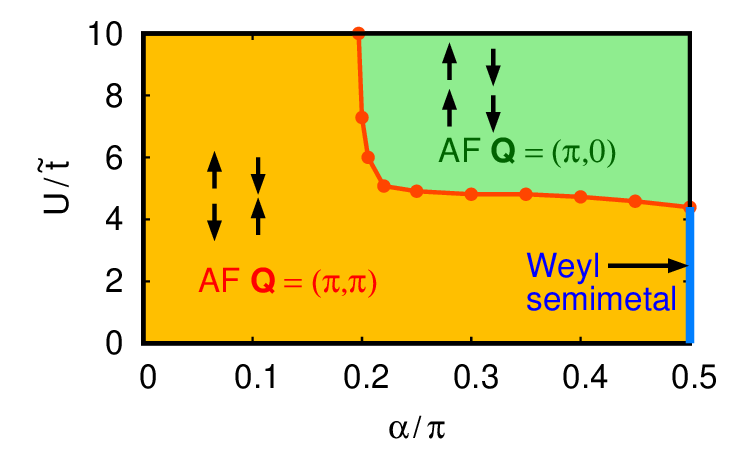}%
  \caption{(Color online)
    Magnetic phase diagram obtained with the Hartree-Fock approximation
    within two-sublattice magnetic order.
    \label{HF_phase_diagram}}
\end{figure}

We have also solved the gap equation for $\mib{Q}=(\pi,0)$
and obtained $\mib{m}_{\text{AF}}$ parallel to the $y$ direction.
By comparing energies for $\mib{Q}=(\pi,\pi)$ and $\mib{Q}=(\pi,0)$,
we construct a phase diagram shown in Fig.~\ref{HF_phase_diagram}.
As noted, the antiferromagnetic state with $\mib{Q}=(\pi,\pi)$ occurs
at infinitesimally small $U$ except for $\alpha=0.5\pi$.
The Weyl semimetallic state remains for $U/~\tilde{t} \lesssim 4.4$
at $\alpha=0.5\pi$.
The antiferromagnetic state with $\mib{Q}=(\pi,0)$ appears
at large $U$ for $\alpha/\pi \gtrsim 0.2$.

This phase boundary can be understood from the effective Hamiltonian
in the strong coupling limit.
The effective Hamiltonian is derived from
the second-order perturbation theory concerning $t$ and $\lambda_R$
and is given by~\cite{Cocks2012, Radic2012, Minar2013, Gong2015}
\begin{equation}
  H_{\text{eff}}
  =
  \sum_{\mib{r} \mib{a} \mu}
  \left[
  J^{\mu}_{\mib{a}}S_{\mib{r}}^{\mu}S_{\mib{r}+\mib{a}}^{\mu}
  +D_{\mib{a}}^{\mu}(\mib{S}_{\mib{r}} \times \mib{S}_{\mib{r}+\mib{a}})^{\mu}
  \right],
\end{equation}
where
$\mib{a}=\hat{\mib{x}}$ or $\hat{\mib{y}}$,
$\mu=x$, $y$, or $z$,
$\mib{S}_{\mib{r}}$ is the spin operator at site $\mib{r}$,
$J_{\hat{\mib{x}}}^x
=J_{\hat{\mib{x}}}^z
=J_{\hat{\mib{y}}}^y
=J_{\hat{\mib{y}}}^z
= 4(t^2-\lambda_R^2)/U$,
$J_{\hat{\mib{x}}}^y
=J_{\hat{\mib{y}}}^x
= 4(t^2+\lambda_R^2)/U$,
$D_{\hat{\mib{x}}}^y
=-D_{\hat{\mib{y}}}^x
=8t\lambda_R/U$,
and the other components of $\mib{D}_{\mib{a}}$ are zero.
From the anisotropy in the interaction,
we expect the ordered moments
along the $x$ or $y$ direction for $\mib{Q}=(\pi,\pi)$
and
along the $y$ direction for $\mib{Q}=(\pi,0)$.
Thus, the directions of the ordered moments
obtained with the Hartree-Fock approximation
are in accord with the effective Hamiltonian.
For $t \ll \lambda_R$ ($\alpha \simeq 0$),
the magnetic order with $\mib{Q}=(\pi,\pi)$ is stable
as in the ordinary Heisenberg model.
For $t \gg \lambda_R$ ($\alpha \simeq 0.5\pi$),
the magnetic order with $\mib{Q}=(\pi,0)$ has lower energy
than that with $\mib{Q}=(\pi,\pi)$ due to the anisotropic interaction.
For $t=\lambda_R$
($J_{\hat{\mib{x}}}^x
=J_{\hat{\mib{x}}}^z
=J_{\hat{\mib{y}}}^y
=J_{\hat{\mib{y}}}^z=0$),
if we ignore the Dzyaloshinskii-Moriya interaction $\mib{D}_{\mib{a}}$,
the model is reduced to the compass model~\cite{Kugel1982}.
It is known as a highly frustrated model.
The condition $t=\lambda_R$
corresponds to $\alpha=\tan^{-1}(1/\sqrt{2})=0.1959\pi$.
Thus, the phase boundary $\alpha \simeq 0.2\pi$
obtained with the Hartree-Fock approximation at a large-$U$ region
is corresponding to the highly frustrated region of the model.

However, in a large-$U$ region, we expect longer-period magnetic order
due to the Dzyaloshinskii-Moriya interaction.
It is out of the scope of the present study
and has already been investigated by previous studies
using the effective Hamiltonian~\cite{Cocks2012, Radic2012, Gong2015}.
Our important finding in this section
is the absence of the paramagnetic phase except for $\alpha=0.5\pi$
in the weak-coupling region.
However, the ordered moment and the energy gain of the antiferromagnetic state
in the weak-coupling region are exponentially small.
Thus, the transition temperature should be very low,
and the effects of this magnetic order should be weak even at zero temperature.
In addition, this magnetic order would be easily destroyed
by perturbations such as the next-nearest-neighbor hopping
breaking the nesting condition~\cite{Beyer2023}.
Thus, the discussions in the previous sections
without considering magnetic order are still meaningful.

\section{Summary}
We have investigated the Rashba-Hubbard model on a square lattice.
The Rashba spin-orbit coupling generates the two-dimensional Weyl points,
which are characterized by non-zero winding numbers.
We have investigated lattices with edges
and found zero-energy states on a lattice with zigzag edges.
The zero-energy states are localized around the edges
and have a helical character.
The large density of states due to the flat zero-energy band
may result in magnetic polarization at edges,
similar to graphene~\cite{Fujita1996}.

We have also examined the effects of the Coulomb interaction $U$.
The Coulomb interaction renormalizes the ratio of
the coupling constant of the Rashba spin-orbit coupling $\lambda_R$
to the hopping integral $t$ effectively.
As a result, the Weyl points can move to the Fermi level
by the correlation effects.
Thus, the Coulomb interaction
can enhance the effects of the Weyl points
and assist in observing phenomena originating from the Weyl points
even if the bare Rashba spin-orbit coupling is not large.

We have also investigated the magnetism of the model
by the Hartree-Fock approximation.
We have found that the antiferromagnetic state with the ordering vector
$\mib{Q}=(\pi,\pi)$ occurs at infinitesimally small $U$
due to the perfect nesting of the Fermi surface
even for a finite $\lambda_R$.
However, the density of states at the Fermi level
becomes small for a large $\lambda_R$
and as a result,
the energy gain by the antiferromagnetic order
is small in the weak-coupling region.
Therefore, the transition temperature is very low,
and the effects of the magnetic order should be weak in such a region.
In addition, this magnetic order would be unstable against perturbations,
such as the inclusion of next-nearest-neighbor hopping~\cite{Beyer2023}.
Thus, we conclude that the discussions on the Weyl semimetal
without assuming magnetism are still meaningful.

\begin{acknowledgment}
This work was supported by JSPS KAKENHI Grant Number
JP23K03330. 
\end{acknowledgment}


\end{document}